\documentclass[]{acm_proc_article-sp}

\usepackage{hyperref}
\usepackage{subfigure}

\hypersetup{pageanchor=false}
\usepackage{amsmath}
\usepackage{hhline}
\usepackage{pbox}
\usepackage{listings}
\newcommand{\urlwofont}[1]
{
\urlstyle{same}\url{#1}
}

\graphicspath{
{fig/}
}

\begin{document}

\title{LODE: Linking Digital Humanities Content\\ to the Web of Data }

\numberofauthors{6} 
\author{
\alignauthor
Jakob Huber\\
       \affaddr{Data and Web Science Group}\\
       \affaddr{University of Mannheim}\\
       %\email{jahuber@mail.uni-mannheim.de}
\alignauthor
Timo Sztyler\\
       \affaddr{Data and Web Science Group}\\
       \affaddr{University of Mannheim}\\
       %\email{tsztyler@mail.uni-mannheim.de}
\alignauthor
Jan Noessner\\
       \affaddr{Data and Web Science Group}\\
       \affaddr{University of Mannheim}\\
       %\email{jan@informatik.uni-mannheim.de}
\and
\alignauthor
Jaimie Murdock\\
       \affaddr{School of Informatics and Computing \& Program in Cognitive Science}\\
       \affaddr{Indiana University}\\
      % \email{jammurdo@indiana.edu}
\alignauthor
Colin Allen\\
       \affaddr{Department of History and Philosophy of Science \& Program in Cognitive Science}\\
       \affaddr{Indiana University}\\
      % \email{colallen@indiana.edu}
\alignauthor
Mathias Niepert\\
       \affaddr{Department of Computer Science and Engineering}\\
       \affaddr{University of Washington}\\
       %\email{mniepert@cs.washington.edu}
}
\date{}
% Just remember to make sure that the TOTAL number of authors
% is the number that will appear on the first page PLUS the
% number that will appear in the \additionalauthors section.

\maketitle

\begin{abstract}
Numerous digital humanities projects maintain their data collections in the form of text, images, and metadata. While data may be stored in many formats, from plain text to XML to relational databases, the use of the resource description framework (RDF) as a standardized representation  has gained considerable traction during the last five years. Almost every digital humanities meeting has at least one session concerned with the topic of digital humanities, RDF, and linked data.

While most existing work in linked data has focused on improving algorithms for entity matching, the aim of the \textsc{LinkedHumanities} project is to build digital humanities tools that work ``out of the box," enabling their use by humanities scholars, computer scientists, librarians, and information scientists alike.

With this paper, we report on the Linked Open Data Enhancer (\textsc{Lode}) framework developed as part of the \textsc{LinkedHumanities} project.
With \textsc{Lode} we support non-technical users to enrich a local RDF repository with high-quality data from the Linked Open Data cloud. \textsc{Lode} links and enhances the local RDF repository without compromising the quality of the data. In particular, \textsc{Lode} supports the user in the enhancement and linking process by providing intuitive user-interfaces and by suggesting high-quality linking candidates using tailored matching algorithms. We hope that the \textsc{Lode} framework will be useful to digital humanities scholars complementing other digital humanities tools.
\end{abstract}
\newpage
\category{H.3.3}{Information Storage and Retrieval}{Information Search and Retrieval}
\category{H.3.7}{Information Storage and Retrieval}{Digital Libraries}[Systems issues]

\terms{Algorithms, Human Factors, Design}

%\keywords{ACM proceedings, \LaTeX, text tagging} % NOT required for Proceedings

\section{Introduction}
A preeminent scholarly problem is how to comprehend the explosion of high-quality scholarship available in digital formats on the Internet. Humanities scholars, like all academics, are increasingly reliant on the World Wide Web for access to scholarly materials and they are rapidly transferring traditional journals and rare archives to digital formats, further exacerbating the problems of information overload.

Every year, digital humanities projects present their work at the International Conference for Digital Scholarship in the Humanities\footnote{\urlwofont{http://dh2013.unl.edu/}} (DH) and the number of collections is growing steadily. The recent introduction of the new track ``digital humanities'' at the Joint Conference on Digital Libraries (JCDL) underlines the importance of this research field.

Available search engines have failed to solve the problem of
meaningful access, and most users, including students and scholars,
lack the necessary skills to construct effective search queries. 
(For an overview of issues relating to the novelty of search querying, see \cite{pollock:1997}.)

In light of these challenges, some digital humanities pro\-jects have begun to build and maintain collections using machine-readable and structured representations such as XML and RDF. In recent years, the Linked Data initiative\footnote{\urlwofont{http://linkeddata.org}} has gained considerable traction.  Its goals are to create large and interconnected collections of open and structured  data repositories. %Examples are (a) VIVO\footnote{\urlwofont{http://vivoweb.org/about}} which provides machine readable facts about researcher interests, activities, and accomplishments, enabling the discovery of research and scholarship across disciplines; and (b) 
Arguably, the most prominent examples is \textsc{DBPedia}~\cite{bizer2009dbpedia}\footnote{\urlwofont{http://dbpedia.org/About}} -- a data repository that contains structured information extracted from Wikipedia. 
In the last few years, Linked Data has become increasingly important in the area of digital humanities. There have been an overabundance of projects like \textsc{JeromeDL}~\cite{kruk2006anatomy},  
\textsc{Talia}~\cite{nucci2007talia} and, more recently, the  Digitized Ma\-nu\-scripts to Europeana (DM2E)\footnote{\urlwofont{http://dm2e.eu/}} project, whose aim is to enable humanities researchers to work with manuscripts in the Linked Open Web. 

The primary motivation for the \textsc{LinkedHumanities} project is that data and knowledge in isolation does not leverage its full potential. By including links between entities within a data collection and to external resources, novel information is created and inferred, making the resulting collection more valuable than the sum of its parts and giving information context and interoperability.
%During the ``Networked Humanities: Art History in the Web''\footnote{\urlwofont{http://www.esf.org/index.php?id=6726}} conference, for instance, participants discussed  how semantic web technology could benefit humanity scholars and merge split collections. 

\begin{figure}[t!]
\begin{center}
\includegraphics[width=0.48\textwidth]{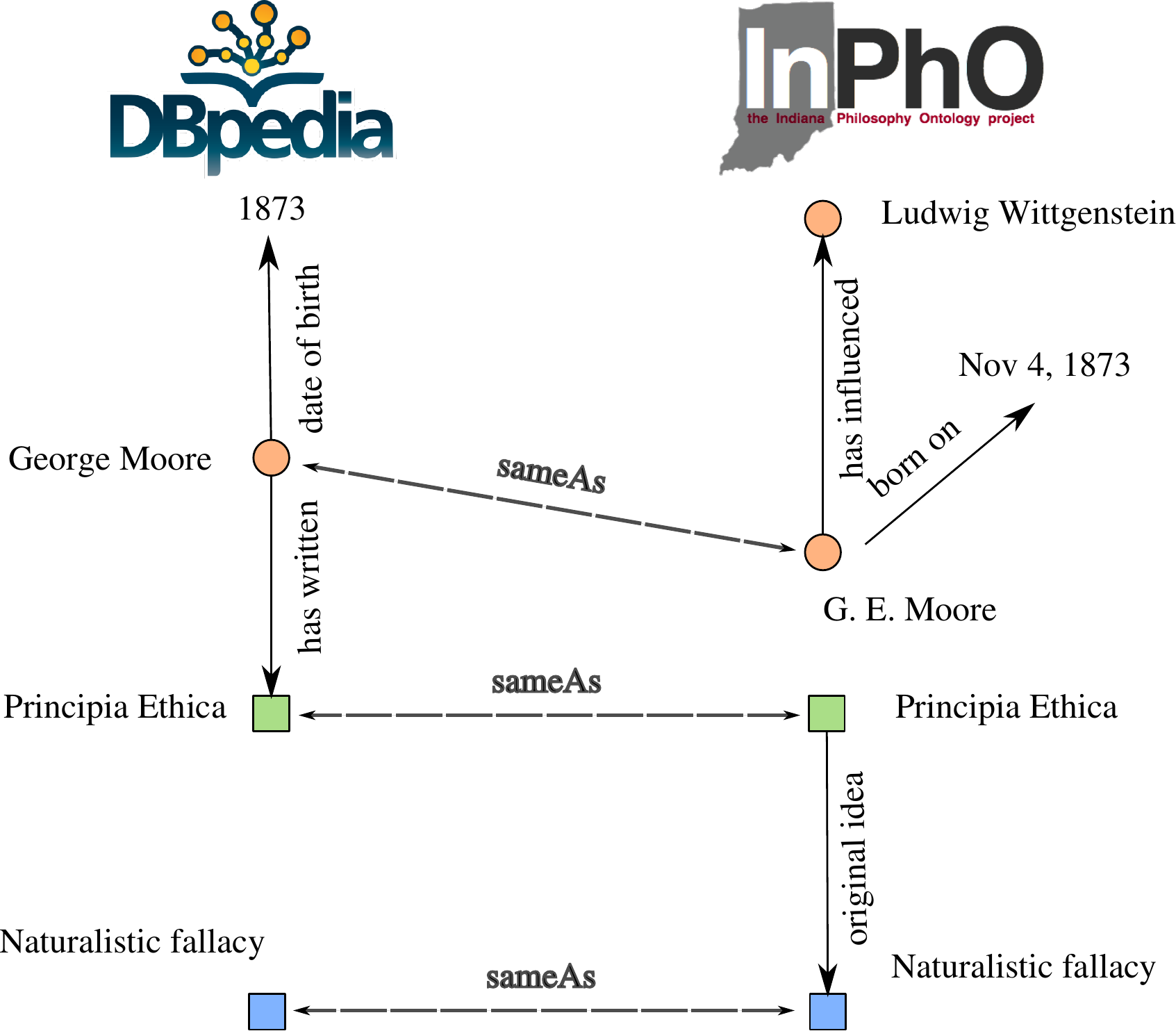}
\caption{Graphical representation of a fragment of RDF data. The most common semantic links between repositories (here \textsc{DBpedia} and \textsc{InPhO}) are \textit{sameAs} links that assert equality between entities.}
\label{fig:rdf-example}
\end{center}
\end{figure}

%The motivation of our approach is the  accumulation of knowledge and the enrichment of currently isolated digital humanities collections by integrating them with the linked data cloud. 
With this paper, we present the linked open data enhancer (\textsc{Lode}) framework, which is the main result of the bi-lateral \textsc{LinkedHumanities} project with the goal to
create and maintain data exploration and integration tools tailored to digital humanities collections so as to help build a machine-readable web of humanities data~\cite{weboffacts2,weboffacts1,dbpedia:2007}. 
%The tool will interconnect structured digital representations of the humanities and leverage the created links to enrich the respective data repositories.
\textsc{Lode} features (a) an explorer component that allows digital humanists to browse and explore local RDF repositories; (b) a linking components that facilitates the linking of local RDF repositories to external RDF repositories such as \textsc{DBPedia}; and (c) an enhancement component for populating and extending the local RDF repository by exploiting the previously created links. In particular, we target use cases that emphasize high quality data requiring human supervision. The linking components provides two different candidate ranking algorithms, which are both suitable for high quality interactive matching candidate selection. We evaluate these linking algorithms with respect to typical digital humanities entities such as documents, concepts, and persons.

\textsc{Lode}'s data integration and enhancement component was designed so as to serve the needs of typical digital humanities projects. In fact, existing projects that maintain RDF collection can benefit from \textsc{Lode}. 
We have explored and report on concrete use cases such as the Indiana Philosophy Ontology (\textsc{InPhO})\cite{Niepert2007,Niepert2008,Buckner2010}\footnote{\urlwofont{http://InPhO.cogs.indiana.edu}} project and the Stanford Encyclopedia of Philosophy (SEP). Furthermore, we show that typical information extraction projects such as the Never Ending Language Learning (\textsc{Nell})\cite{DBLP:conf/aaai/CarlsonBKSHM10}\footnote{\urlwofont{http://rtw.ml.cmu.edu/rtw/}} project, can be used with \textsc{Lode}. \textsc{Nell}  applies machine learning algorithms to continuously extract knowledge from the web and has already accumulated more than $50$ million facts in form of object-predicate-subject triples.

The \textsc{Lode} framework is a collection of integrated digital humanities tools working ``out of the box'' and shielding most of the technological standards and intricacies from its users. The major assumption is that digital humanities projects are run by humanists and librarians not computer scientists. Indeed, most existing work has focused on improving specialized algorithms for entity linking and ontology matching~\cite{euzenat:2011}. While the \textsc{Lode} approach does take advantage of such technologies, the main focus is on user-friendly interfaces. It satisfies all of the criteria for digital humanities infrastructure, as outlined by \cite{crane:2007}, including: 1) named entities (via URIs); 2) a cataloging service (via the RDF relations); 3) structured user contributions (via the linking interface); 4) custom, personalized data (via tools powered by the open-access querying interfaces). The design also follows the guidelines of Stollberg et al.\cite{stollberg:2004}, by providing a concrete example of a semantic web portal.

\section{Linked Open Data}

The term \emph{linked data} describes an assortment of best practices for publishing, sharing, and connecting structured data and knowledge over the web \cite{LDsofar}. These standards include the assignment of URIs to each datum, the use of the HTTP protocol, RDF data model (Resource Description Framework), and hyperlinks to other URIs \cite{Berners-Lee2006}. Whereas the traditional World Wide Web is a collection of documents and hyperlinks between these documents, the data web extends this to a collection of arbitrary objects (resources) and their properties and relations. For example, rather than containing article content, \textsc{DBpedia} represents each \textsc{Wikipedia} article as its own entity, and leverages the link structure between articles as well as structured ``infobox'' data to establish semantic relations \cite{dbpedia:2007}. %If any structured ``infobox'' data are present, such as information for a city indicating the coordinates, country, population, and size, the literal strings and links to other articles contained in each field can be marked with more precise semantic relations.

These relations are modeled using the resource description framework (RDF)\cite{w3:rdf}, a generic graph-based data model for describing objects and their relationships with each other. Further semantic relations like \texttt{broader}, \texttt{narrower}, and \texttt{disjoint with} have been standardized in the Simple Knowledge Organization System (SKOS)~\cite{baker2013key} and in the Web Ontology Language (OWL 2)\footnote{\urlwofont{http://www.w3.org/TR/owl2-overview/}}. We forward the reader to Table~\ref{tab:relations} for a very small subset of such relations. 

Figure~\ref{fig:rdf-example} depicts a small fragment of an RDF graph modeling the domain of philosophers and philosophical works and ideas. Each node in the graph represents a particular object and carries a unique identifier. Links between entities in different data repositories establish semantic connections. For instance, the object with the label ``George Moore'' in \textsc{DBpedia} is identical to the object with the label ``G.E. Moore'' in the Indiana Philosophy ontology which is expressed by the \texttt{sameAs} link between the two entities. Using RDF representation allows us to not only resolve ambiguities but also to \emph{aggregate} knowledge about individual entities. 
%There are many different representation formats, known as serializations, for RDF data. Examples of RDF serializations include RDF/XML, Notation-3 (N3), Turtle, N-Triples, RDFa, and RDF/JSON.

\begin{figure}[t!]
\begin{center}
\includegraphics[width=0.48\textwidth]{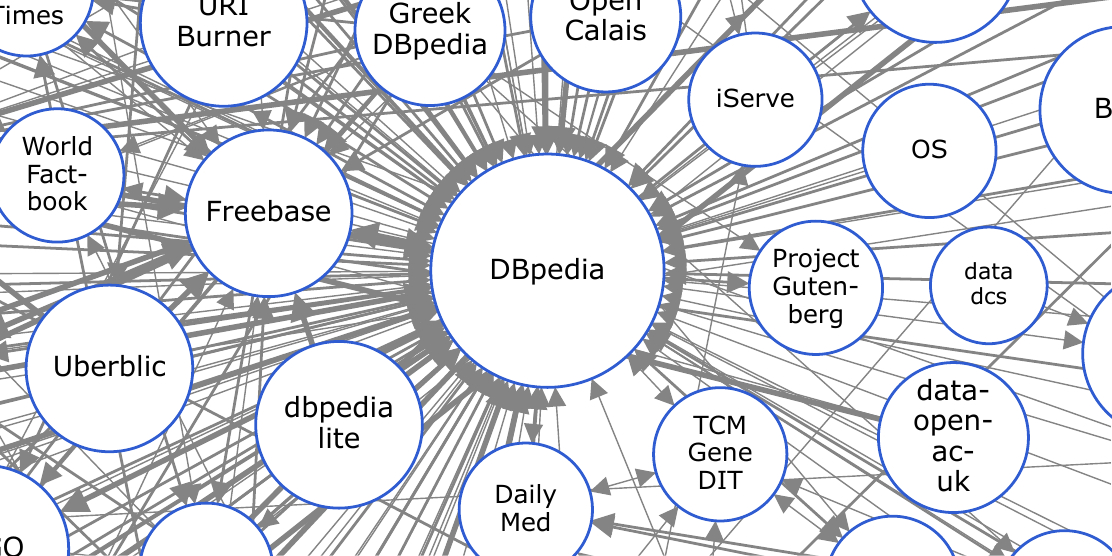}
\caption{A small fragment of the web of data. \textsc{DBpedia} is a de-facto hub of the linked open data could.}
\label{fig:lodc}
\end{center}
\end{figure}

When published under open licenses, data sets may join the Linked Open Data cloud, which can help increase awareness of new collections and facilitates integration with other LOD repositories. For  a small fragment of the LOD web see Figure~\ref{fig:lodc}. %Most repositories in the linked open data cloud are published under a creative commons license meaning that others are free to copy, distribute, adapt, and transmit the data as long as they attribute the creator and/or distribute the transformed data under the same or a similar license. %\textsc{DBPedia}, for instance, is available under licenses CC-BY-SA\footnote{\urlwofont{http://creativecommons.org/licenses/by-sa/3.0/legalcode}}. In fact, the creators of linked open data explicitly encourage the reuse and distribution of data. 

\section{Related Work}

Apart from the \textsc{InPhO} project, there have been many attempts to digitize data in humanities by utilizing Semantic Web technologies. 
The system \textsc{Talia}~\cite{nucci2007talia} enables, for example, philosophy scholars to compare manuscripts, search for specific topics in hand-written paragraphs, and link movie files to the topics. \textsc{Talia} employs RDF as underlying representation formalism. \textsc{Talia} has been developed within the \textsc{OAC} project that creates a model and data structure to enable the sharing of scholarly annotations across annotation clients, collections, media types, applications, and architectures~\cite{nucci2008semantic}. One of the project's goals is the generalization of the tools developed specifically for the philosopher Nietzsche in the Hyper-Nietzsche project~\cite{d2007nietzsche}. 

Additionally, there are several projects which focus on providing a Semantic Web architecture for specific fields. For instance, the \textsc{Bricks} project maintains a service-oriented infrastructure to share knowledge and resources in the Cultural Heritage domain with RDF \cite{risse2005bricks} while the \textsc{JeromeDL} project designed an architecture for social semantic digital libraries~\cite{kruk2006anatomy}. 

More recently, the Digitized Ma\-nu\-scripts to Europeana \\(DM2E)\footnote{\urlwofont{http://dm2e.eu/}} project developed tools which enable humanities researchers to interact with the Semantic Web. If we compare DM2E with our linked humanities project, DM2E allows users to annotate digital humanities collections with existing vocabularies such as SKOS while \textsc{Lode} allows users to use their own project-specific RDF representation (like e.g. in the InPhO project) and provides a framework for browsing, linking, and enhancing this representation. As such, the projects complement one another and we will continue to explore possible synergies of the two projects. 
One of DM2E's objectives is to parse manuscripts and make their data available in Europeana~\cite{haslhofer2011data}, which is a multi-lingual online collection of millions of digitized items from European museums. In this context, DM2E members identified, for instance, some challenges for building an ontology about the philosopher Ludwig Wittgenstein by pointing out different ontological concepts and modeling alternatives~\cite{pichler2013sharing}. M{\'a}cha et al.~\cite{macha2013overlapping} extends this work by approaching the problem of modeling agreement, differences, and disagreement. Another important aspect of DM2E is provenance tracking~\cite{eckert2013Provenance,eckert2014} which makes it possible to identify the sources of linked data with the help of a consistent data model. In this context DM2E developed the tool \textsc{Pundit}~\cite{grassi2013pundit} which allows the annotation, augmentation, contextualization, and externalization of Web Resources of manuscripts so as to make these manuscripts available as machine-processable data. 

This large amount of projects with the aim of providing tools for the creation or maintenance of small and specialized digital humanities datasets, underlines the need to establish high-quality links between these datasets and the LOD cloud. These links can then be utilized for controlled enrichment of datasets with further information.

With respect to the problem of entity matching, a large amount of automated and semi-automated instance matching frameworks have emerged. 
In the context of the DM2E project, the linking framework \textsc{Silk}~\cite{isele2010silk} is utilized for establishing links between data items within different Linked Data sources. It has been integrated into the workflow system \textsc{OmNom}~\cite{gradmann2013modellierung}. In \textsc{OmNom} the user can for example parse different file formats and upload them to the DM2E triple store. Thereby, the input and output parameters of the work-flow steps are intuitively connected via drag-and-drop. Furthermore, each work-flow can be executed by creating a specific instance of the work-flow with concrete input and output files. 
Within \textsc{Silk} the user can define work-flows in form of a tree structure, which describe how the interlinking process is performed. Creating these work-flows, however, requires a significant knowledge about RDF and about specific linking techniques such as different syntactic similarity measures. To that end, \textsc{Silk} provides an approach to actively learn the work-flows by forcing the user to accept and decline a number of matching candidates~\cite{isele2013active}. The \textsc{Lode} framework aims to hide even more complexity from the user, by providing the user valuable suggestions and concise information about the respective candidate entities, without the need to learn a linking scheme. Consequently, we enable domain experts, who have very limited knowledge of Semantic Web technologies to link and enrich their data by using simple drag-and-drop techniques. Furthermore, our use case is such that the user needs to have full control over the link creation process in order to preserve the quality of the local repository. Thus, any automatic link establishment techniques without the supervision of the user is not suitable for our use case.

In addition to \textsc{Silk}, many other tools incorporate the idea of active or supervised learning for matching. 
Existing supervised approaches are based on learning linear classifiers or learning threshold based Boolean classifiers~\cite{arasu2011active,isele2013active}. An example for the first category is \textsc{Marlin} (Multiple Adaptive Record Linkage with Induction)~\cite{bilenko2003adaptive} which uses support vector machines for learning. Examples for the second category include \textsc{Active Atlas}~\cite{tejada2002learning} and TAILOR~\cite{elfeky2002tailor} which utilize decision trees for record linkage. 
In active learning, Arasu et al.~\cite{arasu2010active} developed an approach with which a certain precision can be reached. However, recall might still be low after learning. The tool \textsc{Raven}~\cite{ngomo2011raven} focuses on Boolean and weighted classifiers. Genetic algorithms~\cite{de2010active,ngomo2012eagle,isele2013active} are a common technique for finding solutions for active learning approaches. There also exist several fully automated non-supervised instance matching systems like \textsc{LogMap}~\cite{jimenez2012large}, \textsc{Codi}~\cite{noessnerCODI2010,huber2011codi,noessner2010leveraging}, \textsc{Rimom}~\cite{li2009rimom}. Moreover, there is a large body of work on approaches that exploit schema information to make the resulting alignments more coherent~\cite{meilicke2011alignment,niepert2010probabilistic,Niepert11c,jimenez2012large,noessner2011coherent,niepert2010uai}. While these systems do not require any additional user input they often rely on an expressive ontology for the alignment process. However, these systems' alignments are often error-prone~\cite{noessner2011interactive} and thus not suitable for establishing links of very high quality. We refer the reader to the instance matching track of the ontology alignment evaluation initiative (OAEI) for a listing of further systems~\cite{aguirre2012results}. Again, the focus of the \textsc{Lode} project is the combination of linking algorithms that do not require learning and an expressive schema with intuitive user interfaces for semi-automatic alignment tasks. Furthermore, \textsc{Lode} requires the linking algorithms to be interactive, responding to a query within less than a second. Moreover, since the link creation is semi-automatic, a ranking of possible entities is required.

\begin{figure}[t!]
\begin{center}
\includegraphics[width=0.48\textwidth]{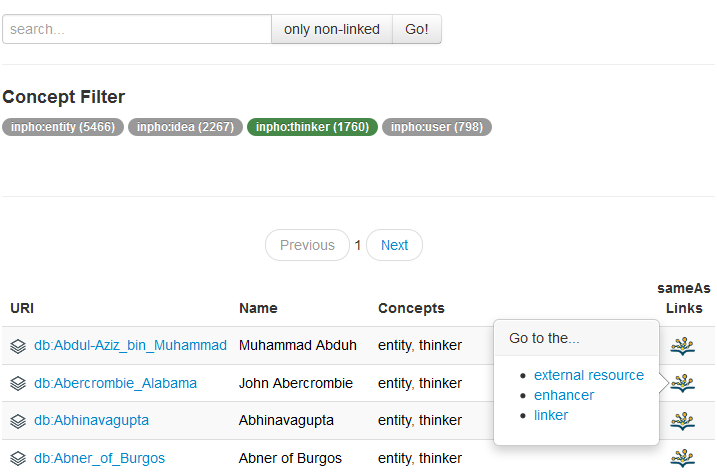}
\caption{Screenshot of the browser interface that lists all search results for ``InPhO:thinker''. The concept filters can be used to focus on particular objects types, here the \textsc{InPhO} types ``entity'', ``thinker'', and ``user''.}
\label{fig:browse}
\end{center}
\end{figure}

\section{The LODE Framework}
\label{sec:lodframework}

The linked open data enhancer (LODE)\footnote{\urlwofont{http://lode.informatik.uni-mannheim.de}} framework is a set of integrated tools that allow digital humanists, librarians, and information scientists to connect their data collections to the linked open data cloud. The initial step is to model the respective collection with some RDF serialization. For this task, tools from e.g. the DM2E project can be utilized. Once an RDF representation exists, the \textsc{LODE} framework loads the RDF representation of the collection and provides several components for browsing, integrating, and enriching the collections. While it leverages state of the art concepts and algorithms, % for entity linking, scalable data management, and statistics 
the focus is on intuitive interfaces that shield the users from the algorithmic intricacies and the complexities of RDF serialization. 

%The use-case that we introduce here to describe the framework is the \textsc{InPhO} project's~\cite{Niepert2007} RDF serialization modeling the domain of philosophy. However, every RDF repository can be loaded within \textsc{LODE}. 

In the following, we describe the three modules of \textsc{LODE} in more detail. 

\subsection{Content Browsing}
\label{ssec:browsing}

\begin{figure}[t!]
\begin{center}
\includegraphics[width=0.48\textwidth]{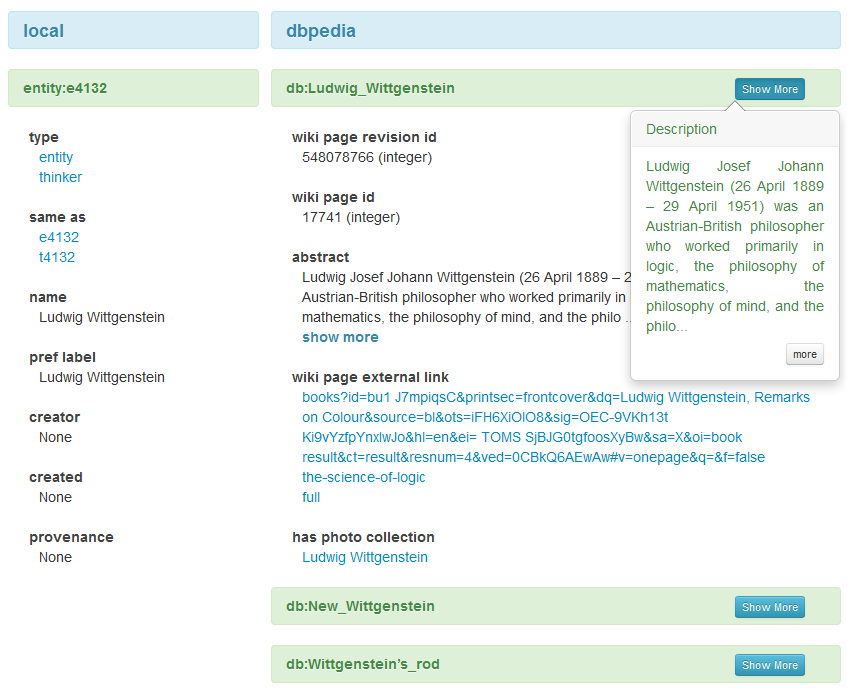}
\caption{Screenshot of the linking interface listing the linking candidates for the \textsc{InPhO} entity ``Ludwig Wittgenstein.'' The link candidate ranking on the right is influenced by the lexical similarity and the semantics of the source entity. Users can easily identify the correct linking candidate by browsing its semantic properties and \textsc{Wikipedia} abstract.}
\label{fig:linking}
\end{center}
\end{figure}

The content browser allows to explore the RDF dataset in an intuitive way by providing a search-based interface that resembles those of standard search engines. Users can enter keywords to search for entities in the locally stored RDF serialization of the project content. All the objects in the RDF dataset that match a given keyword query are categorized according to their types. The search field features auto completion and allows filtering by type. %\textsc{LODE} only displays the types which are entailed in the current search results. 
The syntax for this latter filter technique is adapted from the typical search engine syntax which allows searching for terms within a specific site with the command \textit{site:url searchterm}. We adopted this syntax and applied it to types such that the user can search with e.g. \textit{concept:human Wittgens} for an individual whose label matches the string \textit{Wittgens} and which is an instance of the type \textit{Human}. In addition, the \textsc{Lode} search interfaces provide dynamic faceted search. The results are clustered according to the \textit{sameAs} relations so that every unique entity is only displayed once in the result. Figure~\ref{fig:browse} provides a screenshot of the main functionality of the content browser.

The content browser is the starting point for navigation to the instance overview, linker, and enhancer. The overview displays all data of an instance which are stored in the local triple store. For each local entity, the linking module suggests high-quality linking candidates from the Linked Open Data cloud (here: DBpedia) to the user. If a link exists, the enhancement module gives the user the possibility to decide which information of the LOD cloud is reliable enough to be added to the local RDF repository.

As with every \textsc{LODE} component, the browsing interface works ``out of the box'' for arbitrary RDF repositories and does not need prior configuration steps. %In addition to the properties of the search results, the interface also indicates whether entities are linked to other linked data repositories such as \textsc{DBPedia}.
Like all other pages, the browser provides tooltips which contain the full URI of any displayed instance, concept, or property. The tooltips are clickable and lead the user to the original resource taking advantage of the important linked open data principle that every entity's URI is resolvable.

\begin{table}
	\centering
	\begin{tabular}{|c|c|c|c|c|}
	\hline
	\textbf{Vocabulary}\footnotemark[11] & \textbf{Relation} & \rotatebox{90}{\textbf{Type}} & \rotatebox{90}{\textbf{Usage}\footnotemark[12] }  \\ \hline \hline
	owl & sameAs & I & 34.40\%  \\ \hline
	rdfs & subClassOf & C & 15.28\%  \\ \hline	
	rdfs & subPropertyOf & P & 11.53\%  \\ \hline
	owl & inverseOf & P & 6.65\%  \\ \hline
	skos & broader & I\slash C & 4.87\%  \\ \hline
	owl & equivalentClass & C & 4.40\% \\ \hline
	skos & narrower & I\slash C & 3.84\%  \\ \hline
	owl & disjointWith & C\slash P & 3.56\%  \\ \hline
	owl & equivalentProperty & P & 3.09\%  \\ \hline
	skos & related & I\slash C & 2.81\%  \\ \hline
%	owl & differentFrom & I & 0.01\% \\ \hline
	\end{tabular}
	\caption{Most frequent relation properties in the LOD web.}
	\label{tab:relations}
\end{table}
\footnotetext[11]{Prefixes taken from \urlwofont{http://prefix.cc}}
\footnotetext[12]{The usage has been computed from data representing the LOD web. The data has been crawled by taking seeds from the Billion Triple Challenge \urlwofont{http://km.aifb.kit.edu/projects/btc-2012/} and Datahub \urlwofont{http://datahub.io}.}

\subsection{Content Linking}
\label{ssec:linking}

When an entity is selected in the browsing interface the user can initiate the linking component. At this point, the \textsc{Lode} framework supports linking to \textsc{DBpedia} which is a hub in the link open data cloud (see Figure~\ref{fig:lodc}), providing links to numerous LOD collections through \texttt{sameAs} links. 

Please note that, for the particular applications \textsc{Lode} is aimed at, the user requires full control over the linking process in order to assure the quality standard of the local RDF repository. Hence, the purpose of the linking component is to recommend high quality suggestions from which the user then can select the correct one. Additionally, we especially aim at supporting non-technical domain experts for data integration by designing simple user interfaces and providing them valuable additional information of the linking candidates to facilitate the alignment decision. Figure~\ref{fig:linking} depicts the linking interface for the \textsc{InPhO} entity ``Ludwig Wittgenstein'' and some of the linking candidates. 

In addition to \texttt{sameAs} links, \textsc{Lode} supports different types of links modeling relationships between individuals, typed (concept), and properties. We utilize as subsets of   SKOS~\cite{baker2013key} and also include several relations from the Web Ontology Language (OWL 2). Table~\ref{tab:relations} lists the core link types supported by \textsc{Lode}. We differentiate between relations between concepts (C), properties (P), and individuals (I). This list can be extended by the user at any time.

The content linker performs the following steps to retrieve and display the linking candidates for a candidate entity $\mathsf{E}$ to the user.

First, the linker component extracts a set of search terms from property assertions of entity $\mathsf{E}$ in the local RDF repository. To identify these terms, the algorithm maintains a list of the most frequent properties describing the instance (like e.g. the \texttt{label}). Table~\ref{tab:properties} depicts a list of common lexical properties of entities and their usage statistic. Of course, it is possible to modify and extend this list. However, if the local RDF  repository follows modeling standards common to linked data repositories the list of properties should be sufficient as it covers a large fraction of the properties used for labeling entities.

\begin{table}
	\centering
	\begin{tabular}{|c|c|c|}
	\hline
	\textbf{Vocabulary}\footnotemark[11] & \textbf{Property} & \textbf{Usage}\footnotemark[12] \\ \hline \hline
	foaf & name & 53.14\%  \\ \hline
	rdfs & label & 40.02\% \\ \hline
	foaf & givenname & 21.46\%  \\ \hline
	foaf & accountname & 20.34\%  \\ \hline
	foaf & family\_name & 18.46\%  \\ \hline
	foaf & firstname & 13.96\%  \\ \hline
	foaf & surname & 13.03\%  \\ \hline
	skos & preflabel & 8.62\% \\ \hline
	foaf & openid & 7.50\%  \\ \hline
	dcterms & identifier & 5.81\%  \\ \hline
	\end{tabular}
	\caption{Most frequent label properties in the LOD web.}
	\label{tab:properties}
\end{table}

With the previously extracted search terms as input, the linking component generates a list of potential linking candidates for $\mathsf{E}$ based on two algorithms. 
Both algorithms are required to be interactive, returning a result ranking within one second. Due to common hashing and indexing techniques our algorithms' complexity is sublinear with respect to the total number of possible instances. Section~\ref{ssec:algo1} and Section~\ref{ssec:algo2} provide further details about the linking algorithms.

Finally, \textsc{Lode} extracts context for each linking candidate to help the user identify the correct alignment without overwhelming her with too much information. The context is extracted so as to help the user discriminate between entities with identical labels and names. The underlying selection process is explained in Section~\ref{ssec:selection}. Figure~\ref{fig:linking} shows how  context (abstract, labels, etc.) is presented to the user so as to help the user with the linking decision. 

\subsubsection{SPARQL-Algorithm}
\label{ssec:algo1}

The first linking algorithm uses SPARQL queries to search for matching candidates in the LOD cloud. As an example, we employ \textsc{DBpedia} as \textsc{SPARQL} endpoint. However, please note that we are able to apply the following search technique to any other triple store. 

The SPARQL queries search for the exact search terms within the label and the abstract. Listing~\ref{lst:sparql} provides a simplified example of such a SPARQL query. Especially, when we include the search within the abstract, we obtain a relatively large amount of linking candidates.  

\begin{lstlisting}[captionpos=b, caption=Search by abstract, label=lst:sparql, basicstyle=\ttfamily,frame=none]
PREFIX dbp: <http://dbpedia.org/property/>
SELECT DISTINCT ?instance ?value
WHERE {
 ?instance dbp:abstract ?value .
 ?value <bif:contains> searchterm .
}
\end{lstlisting}

This leads to the requirement to rank the retrieved linking candidates in a second step. For this ranking, we apply the Levenshtein similarity~\cite{levenshtein1966binary} between the search term of the local instance and the linking candidates. If more than one search term exists, the maximum similarity is taken. Intuitively, the higher the similarity, the higher the ranking of the candidate. We used the Levenshtein similarity since it can handle spelling errors like e.g. \textit{Ludwig} and \textit{Ludwik}. 

Within this algorithm, we also consider structural information by leveraging known semantic relationships between types of the involved RDF datasets~\cite{noessner2010leveraging}. A matching candidate is inferred to be disjoint if its types are disjoint with the types of the searched entity.
For instance, if the ``Thinker'' type in \textsc{InPhO} and the ``PhilosophicalTradition'' type in \textsc{DBpedia} are disjoint then the linking interface will exclude all entities of the later type as linking candidates for equivalence links. The disjointness relationships has to be established only once by the user of \textsc{Lode}.

Finally, we apply some \textsc{DBpedia} specific optimizations. In particular, we evaluate whenever the URI of the found instance is a \textit{redirect} or a \textit{disambiguation page} and resolve the URI if this is the case.
% pre-load a predefined number of the found instances and present them to the user. During this step we also
\begin{table}%
\centering
\begin{tabular}{|l|l|c|}
	\hline
	\textbf{Anchor $a$} & \textbf{Simplified URI $u$} & \pbox{20cm}{\textbf{number} \\\textbf{$\#(u,a)$}} \\
	\hline
	\hline
		Plato							& Plato (Philosopher)	& 3560 \\
		PLATO							& PLATO (computer system) &	47 \\
		Plato							& Plato, Missouri		& 20 \\
		Plato							& Plato (crater)		& 15 \\
		Plato							& Beer measurement	& 13 \\
		Plato							& Plato, Magdalena	& 9 \\
		Platon						& Plato (Philosopher)	& 6 \\
	\hline
\end{tabular}
\caption{An excerpt of the search result of the \textsc{WikiStat} table when searching for the anchor texts ``Plato'' and ``Platon''. The entries are ordered by $\#(u,a_n)$, descending.}
\label{fig:WikiStatTable}
\end{table}

\subsubsection{WikiStat Algorithm}
\label{ssec:algo2}

The \textsc{WikiStat} algorithm is based on the idea of exploiting Wikipedia's link structure to compute, for a given search string, the conditional probability of a Wikipedia article given the search string. Consider, for example, the article about philosophy which contains a link to the article with URI http://en.wikipedia.org/wiki/Plato and anchor text ``Plato." This link would increase the conditional probability of the URI given the search string ``Plato." As in Dutta et al.~\cite{duttaintegrating}, we utilize the \textsc{WikiPedia} preprocessor \textsc{WikiPrep} \cite{gabrilovich2006overcoming,gabrilovich2007computing} which computes a table consisting of the anchor-text $a$, the source URI, and the corresponding target URI $u$. We use these tables to compute the conditional probabilities. 
However, as opposed to previous work, we have to incorporate multiple search strings $a_1, a_2, \ldots, a_n$ since an entity can have multiple properties that relate the entity to its label or name. Each search term extracted from the properties in Table~\ref{tab:properties} has to be matched against possible anchors used to link to an article in Wikipedia. 

Let $u$ be a Wikipedia URI and $a_1, \ldots , a_n$ be the extracted search strings. Then, the ranking of the matching candidates is based on the following conditional probability 
$$P(u | a_1 \vee \ldots \vee a_n) = \frac{P(u, a_1 \vee \ldots \vee a_n)}{P(a_1 \vee \ldots \vee a_n)} =$$
$$\frac{\#(u,a_1) / N+\cdots+\#(u,a_n) / N}{\#(a_1) / N+\cdots+\#(a_n) / N},$$
where $\#(u,a_i)$ is the number of $(u,a_i)$ pairs, that is, the number of Wikipedia links to entity $u$ with anchor text $a_i$, $\#(a_i)$ is the number of Wikipedia links with anchor $a_i$, and $N$ is the number of all Wikipedia links. The ranking of the linking candidates is determined by sorting the conditional probability of all URIs $u$ in descending order.

Since we are only interested in the final ranking, we are able to further simplify the above equation. In fact, it is sufficient to compute 
$$\#(u,a_1)+\cdots+\#(u,a_n)$$ 
for every URI $u$ because $\#(a_1)+\cdots+\#(a_n)$ is constant for given anchor texts $a_1, a_2, \ldots, a_n$ and $N$ cancels out. 

For efficiency reasons, we precomputed all numbers $\#(u,a)$ for every URI $u$ and anchor text $a$ and stored these in a relational database table. Table~\ref{fig:WikiStatTable} depicts an excerpt of the table for the anchor texts $a_1=$``Plato'' and $a_2=$``Platon''. 
The ranking of the linking candidates is now computed by selecting every row in the table where the anchor $a$ matches a search string, aggregating the result set with respect to the URI, and sorting the aggregation according to the sum of all numbers $\sum_i \#(u,a_i)$ in descending order. Listing~\ref{lst:sql} shows an example SQL query which again queries for anchor texts $a_1$ = ``Plato'' and $a_2$ = ``Platon''. 

\begin{lstlisting}[captionpos=b, caption={\textsc{WikiStat} SQL query for $a_1=$``Plato'' and $a_2=$``Platon''}, label=lst:sql, basicstyle=\ttfamily,frame=none,mathescape=true]
SELECT $u$, SUM($\#(u,a)$) AS $s$ FROM table 
WHERE $a=$`Plato' or $a=$`Platon' 
GROUP BY $u$ ORDER BY $s$ DESC;
\end{lstlisting}

In this query, the highest ranking is achieved by the simplified URL $u=$``Plato (Philosopher)'' having the quantity $\#(u,a_1) + \#(u,a_2) = 3566$.

\subsubsection{Context Selection}
\label{ssec:selection}

The URI of an entity is often not sufficient for the user to select the correct entity from the set of matching candidates. Even in the presence of labels, choosing the correct entity might be difficult due to ambiguous labels. Therefore, we provide the user with contextual information in form of the entity's properties that more closely characterizes each of the matching candidates and help the user to select the correct entity. Presenting every property of an entity would overwhelm the user with information. Hence, we developed an algorithm that presents \textit{discriminating} properties and types only. After experimenting with alternative, more sophisticated adaptations of TF-IDF, we noticed that the frequency of properties is most helpful in identifying valuable assertions. Thus, we implemented the following algorithm.

Let $p$ be a property and $\mathcal{E}$ be the set of all entities in \textsc{DBpedia}. Furthermore, let $I_p : \mathcal{E} \rightarrow \{0,1\}$ be an indicator function for property $p$ defined as:
$$I_p(\mathsf{E}) = \left\{ \begin{array}{ll}1 \ \ & \text{if $\mathsf{E}$ has as least one assertion for $p$}\\
                                      0     &  \text{otherwise}\end{array}\right\}$$
Then, the frequency $f_p$ for the property $p$ is defined as
$$f_p= \frac{\sum_{\mathsf{E} \in \mathcal{E}} I_p(\mathsf{E})}{|\mathcal{E}|}.$$

The frequencies $f_p$ are precomputed for each property. The properties are now sorted according to their frequency in descending order. Finally, we present only the $k$ most frequent properties of an entity as its context. Analogously to properties, we apply the same approach to the types of an entity presented to the user.

\subsection{Content Enhancing}
\label{sec:enhancing}

\begin{figure}[t!]
\begin{center}
\includegraphics[width=0.48\textwidth]{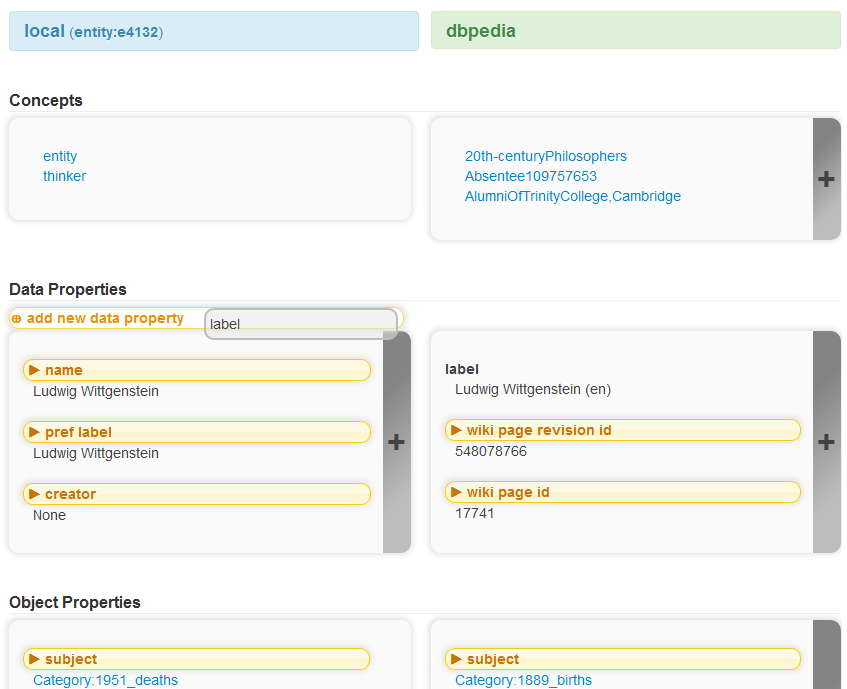}
\caption{Screenshot of the enhancer interface listing the data and meta-data of the local RDF repository (on the left) and related \textsc{DBPedia} content (on the right). Data, concepts, and properties related to the source entity (here: Wittgenstein) can be dragged and dropped to the local repository to enrich the project content.}
\label{fig:enhance}
\end{center}
\end{figure}

After a link between a local and external entity has been established, the enhancing component facilitates the addition of content from the Linked Open Data cloud to the local repository. Since the data of several information extraction projects such as \textsc{DBPedia} contains factual errors and inaccuracies, we allow the user to manually drag and drop LOD content to the local repository. This ensures that the quality of the local collection is not compromised.  The human domain expert verifies the correctness of facts by dragging these facts to the local repository.

The main objective of the component is to support non-technical users with (a) an intuitive interface and (b) high quality enhancement suggestions. 
Figure~\ref{fig:enhance} shows a screenshot of our enhancement interface. The local RDF repository is depicted on the left while \textsc{DBPedia} is located on the right. The interface avoids overwhelming the user with too many potential enhancement candidates by presenting  only excerpts of the most frequent class and property assertions. Here, we utilize the same algorithm as described in Section~\ref{ssec:selection}. 

If the user has decided to enhance a specific class or property value, she can simply drag and drop it to the desired position. During this process, the user gets all possible drop areas highlighted. In Figure~\ref{fig:enhance}, the user decided to enhance the local entity ``Ludwig Wittgenstein'' with a property assertion stating that the entity has a label ``Ludwig Wittgenstein''. 

In case of property assertions, the user has the choice between adding the value to an existing property or creating a new property and assigning the value to this property. If there exists more than one value for a specific property, the user can select which of the given values are to be added to the local collection. %After moving a value from the right to the left side it disappears on the right side. Hence, the user sees only values on the right side which are not already stored for the current individual in the local triple store. 
Additionally, we provide the possibility to delete concepts, properties, or values. %This symbol appears when the user clicks on a local entity. After deletion, the value is displayed on the right side again.

%A subset of these relations are reused to improve the results of the linker interface by excluding linking candidates which are in conflict with each other. Similarly, the relations are also reused by the enhancing interface such that the right side provides only values which are not in conflict with the already stored values. 

Internally, \textsc{LODE} creates new RDF triples in the local RDF repository for each enhancement operation. In our example, \textsc{LODE} will add the new triple ``thinker:t4132 rdfs:label 'Ludwig Wittgenstein'@en'' to the local RDF repository. By keeping the target URI of \textsc{DBPedia} unchanged, it is easily possible to identify the provenance of the enhancement.

\section{Experiments}

Many digital humanities collections are concerned with three different types of entities, namely persons, documents, and concepts. The following experiments assess the performance of the two linking algorithms described in Section~\ref{ssec:linking} on these  different types of entities by using \textsc{InPhO} data. Moreover, we also use a large collection of subject predicate object triples extracted by a well-known information extraction system, namely the Never Ending Language Learning (\textsc{NELL}) project. The experiments are meant to investigate the quality of the real-time interactive linking algorithms with respect to varying types of entities and varying quality of the data.  %In future work, we plan to conduct a user study to especially evaluate the systems usability.
%We measure the performance with the mean reciprocal rank and average runtime results. Furthermore, we examine the influence of different configurations. 

\subsection{Experimental Setup and Datasets}
\label{ssec:eval_approach}
We compare the \textsc{WikiStat} algorithm (see Section~\ref{ssec:algo2}) with the SPARQL algorithm (see Section~\ref{ssec:algo1}) using different configurations. In particular, we considered the properties \texttt{rdfs:label} (abbr. \textit{L}) and \texttt{rdfs:abstract} (abbr. \textit{A}) for the SPARQl queries. Thus, we obtain four different algorithms, which are depicted in Table~\ref{tab:eval_config}.

\begin{table}%
	\centering
	\begin{tabular}{|l|cc|}
				\hline
				\textbf{algorithm} & \textbf{abstract} & \textbf{label}\\ \hline
				SPARQL-A		 & \checkmark 	&	\\ 
				SPARQL-L		 &				      & \checkmark \\ 
				SPARQL-AL	 & \checkmark	  & \checkmark \\ 				
				WikiStat		 &	-						& - \\ \hline
			\end{tabular}
		\caption{Configurations of the \textsc{SPARQL} and \textsc{WikiStat} algorithms (``A'' = abstract considered, ``L'' = label considered)}
		\label{tab:eval_config}
\end{table}

\begin{table}%
\centering
\begin{tabular}{l|c|c|}
		\cline{2-3}
						& \textbf{evaluated} & \textbf{total} \\
		\hhline{~==} \cline{1-1}
		\multicolumn{1}{|l|}{\textsc{InPhO} thinker}	&	1452 & 1758 \\ \hline
		\multicolumn{1}{|l|}{\textsc{InPhO} journal}	&	219  & 1122 \\ \hline
		\multicolumn{1}{|l|}{\textsc{InPhO} idea}		&	236  & 2322 \\ \hline
		\multicolumn{1}{|l|}{NELL}		&	921  & $\approx$ 2 Mio.\\ \hline
\end{tabular}
\caption{Number of entities of the gold standard compared to the the total number entities in the datasets.}
\label{fig:InPhO_entities}
\end{table}

For each of the four methods we take a set of $N$ entities for which a \texttt{owl:sameAs} relation to \textsc{DBpedia} exists. For each of these entities, we compute $10$ matching candidates in form of a ranked list. In order to assess the accuracy of the algorithm we compute the average mean reciprocal rank (MRR). For each entity the linking algorithms generate a ranking of which at most one entry is the correct one.  The MRR of a number of rankings is defined as
\begin{equation}
MRR = \frac{1}{|N|} \sum_{i=1}^{|N|} \frac{1}{rank_i},
\label{eq:mrr}
\end{equation}
where $rank_i$ represents the position of correct entity in the returned ranking. By standard convention, we set $\frac{1}{rank_i} = 0$ if the correct entity is not in the ranking. 
In addition to the MRR, we also measure the average time needed to compute the ranking for one entity. All experiments were executed on a virtual machine running on a two core Intel Xeon 4C E5-2609 80W processor with 2 GB of RAM. %The \textsc{DBPedia} triple store operates on the same server but on another virtual machine with 8GB RAM since also other services access this triple store.

%\subsection{Datasets}
%\label{ssec:eval_datasets}

For our evaluation we created gold standards using data from the \textsc{InPhO} and \textsc{Nell} projects. Both projects provide a large collection of subject-predicate-object triples with \textsc{InPhO} focusing on the domain of philosophy and \textsc{Nell} being more focused on popular domains such as sports and movies. For \textsc{Nell} we used an existing gold standard\footnote{\urlwofont{https://madata.bib.uni-mannheim.de/65/}} which provides \texttt{owl:sameAs} links to \textsc{DBpedia} entities for the subject and the object for $1200$  \textsc{NELL} triples.

The remaining three gold standard data sets were extracted from the Indiana Philosophy Ontology \textsc{InPhO}\footnote{\urlwofont{https://InPhO.cogs.indiana.edu/}} project. Like in many humanities domains, the data in the \textsc{InPhO} project mainly describes entities representing persons (\texttt{Think\-ers}), documents (\texttt{Journals}), philosophical concepts (\texttt{Ideas}), and their relations. For each of these categories, we manually created separate gold standards. 

\begin{figure}[t]
	\centering
	\includegraphics[width=0.48\textwidth]{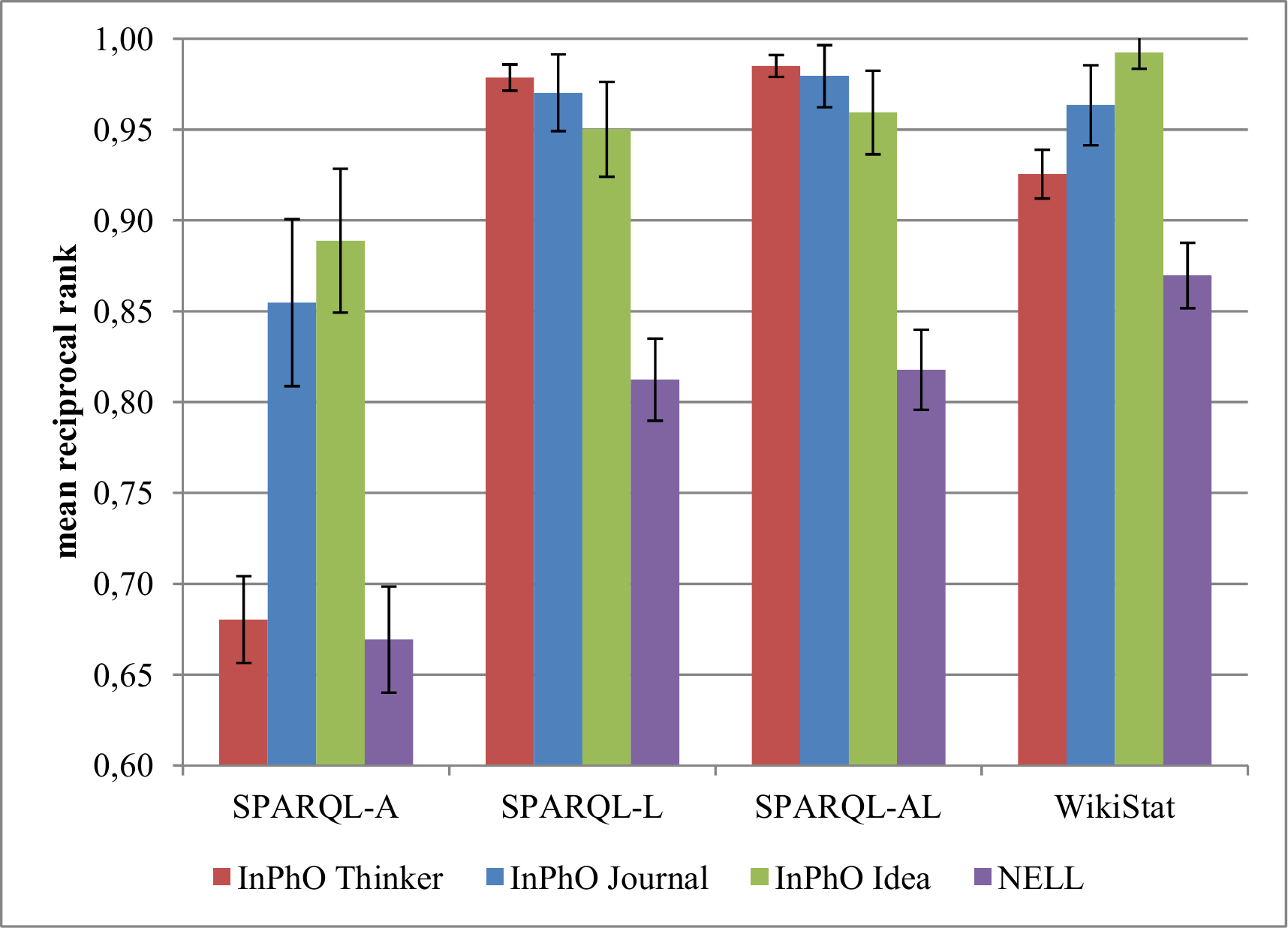}
	\caption{MRR for the different configurations of the search algorithms including the respective 95\% confidence intervals. With respect to SPARQL, we observe that considering the label is crucial. The SPARQL-Algorithm is stronger on thinkers and journals while \textsc{WikiStat} has better results for ideas and on the \textsc{NELL} benchmark.}
	\label{fig:eval1_mrr}
\end{figure}

%In the following, we refer to the datasets as \textsc{NELL} and \textsc{InPhO} datasets.
Table~\ref{fig:InPhO_entities} illustrates the number of individuals per benchmark and category for the gold standards compared to the total available number of entities.

%\begin{figure}[t!]
%\begin{center}
%	\includegraphics[width=0.45\textwidth]{lode_eval_searchterms}
%	\caption{The bar char shows how many search terms, e.g. labels or names, are given per entity and how many can be added by our preprocessing algorithm.}
%	\label{fig:eval1_searchterms}
%\end{center}
%\end{figure}

\subsection{Results}\label{ssec:eval_results}

\begin{figure}[t]
	\centering
	\includegraphics[width=0.48\textwidth]{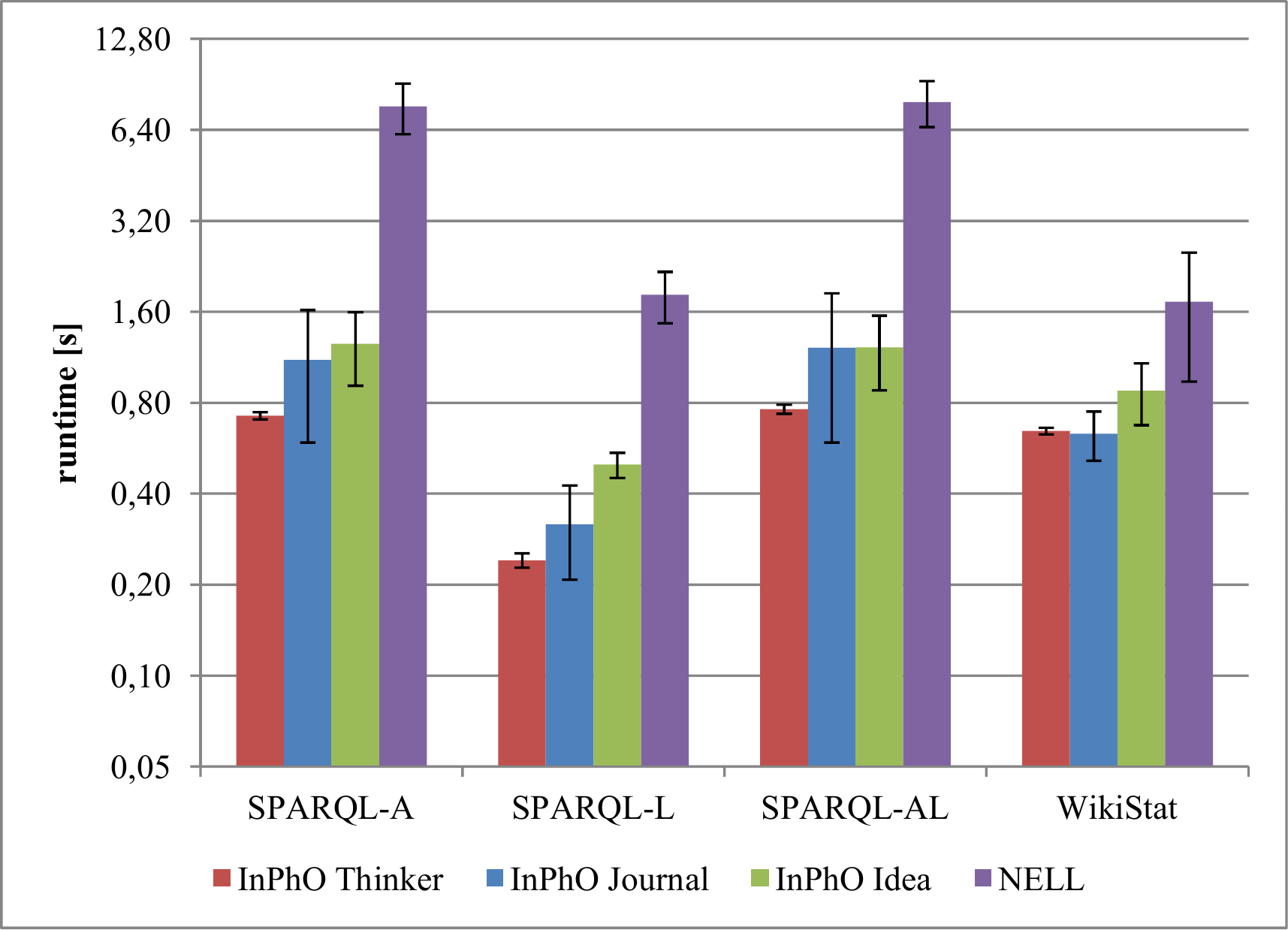}
	\caption{Average runtime for the different configurations of the search algorithms including the respective 95\% confidence intervals in logarithmic scale. The lowest average runtimes are measured for the \textsc{SPARQL}-L configuration. Furthermore, average runtimes were longer for the \textsc{NELL} benchmark.}
	\label{fig:eval1_runtime}
\end{figure}

The MRR values and the average running time with corresponding 95\% confidence intervals are depicted in Figure~\ref{fig:eval1_mrr} and Figure~\ref{fig:eval1_runtime}, respectively. Each figure has four groups of bars  representing the four different configurations depicted in Table~\ref{tab:eval_config}. 
Overall, the linking algorithms achieve MRR values of over 0.95 with the \textsc{SPARQL}-AL configuration on all \textsc{InPhO} entity types with average running times of $1.5$ seconds or less. On the \textsc{NELL} benchmark, the \textsc{WikiStat} algorithm has a MMR over 0.85 and average running times of $1.7$ seconds.

If we compare the MMR results for the different configurations of the \textsc{SPARQL}-based algorithm, we see that considering only the abstract (abbr. \textit{A}) results in the lowest MRR results. Considering the abstract and the label (abbr. \textit{AL}) produces slightly better results than if we only consider the label (abbr. \textit{L}). However, recall can be slightly increased if the label is also considered.
Searching in the abstract only resulted in about the same running times as searching in both the label and the abstract. Considering the label only resulted in the shortest running times which is due to the length of a label being  shorter than the length of an abstract in most cases. 

\begin{figure*}[t]
	\centering
	\subfigure[\textsc{InPhO} Thinker.]{\includegraphics[width=0.24\textwidth]{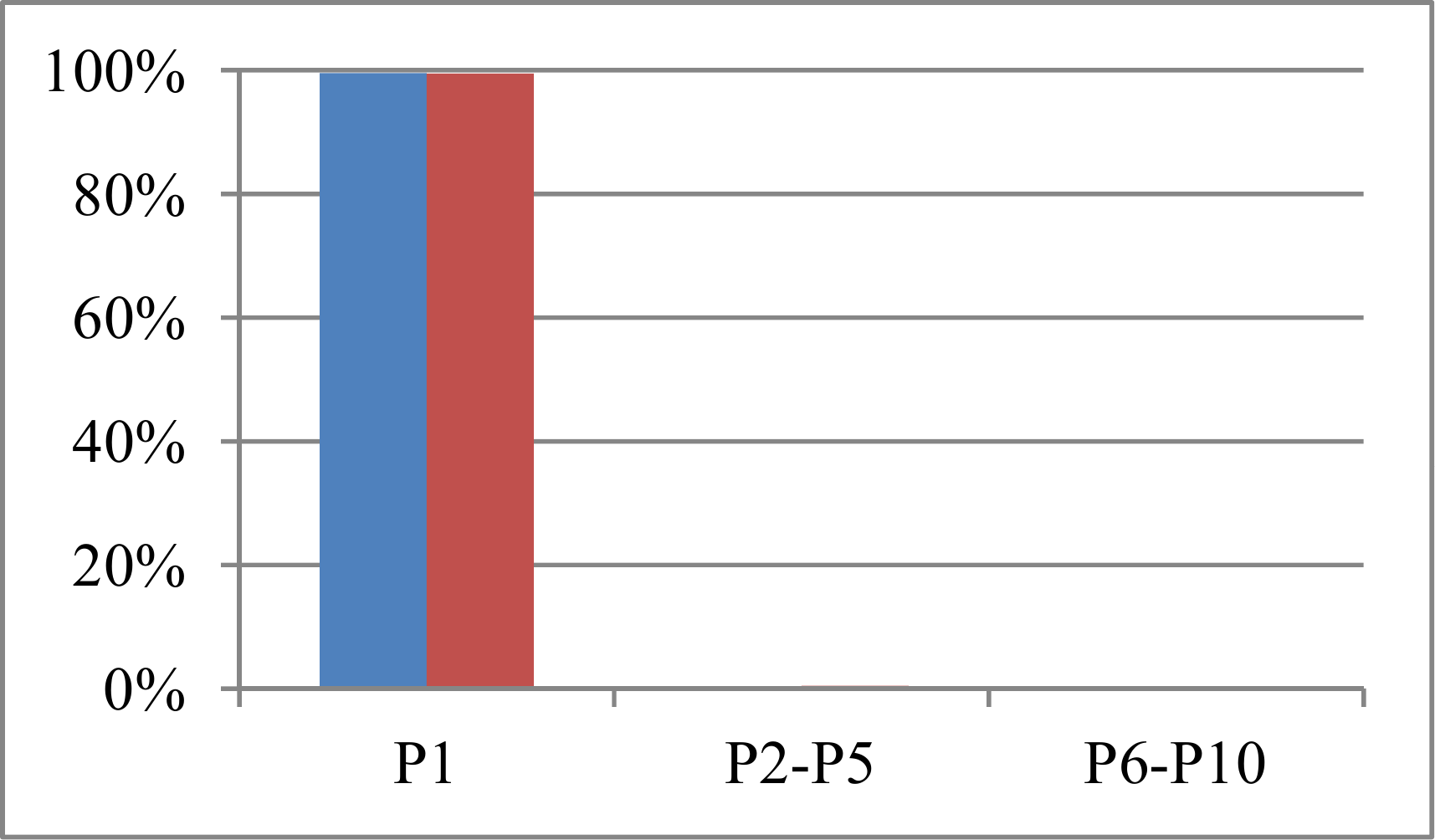}}
	\subfigure[\textsc{InPhO} Journal.]{\includegraphics[width=0.24\textwidth]{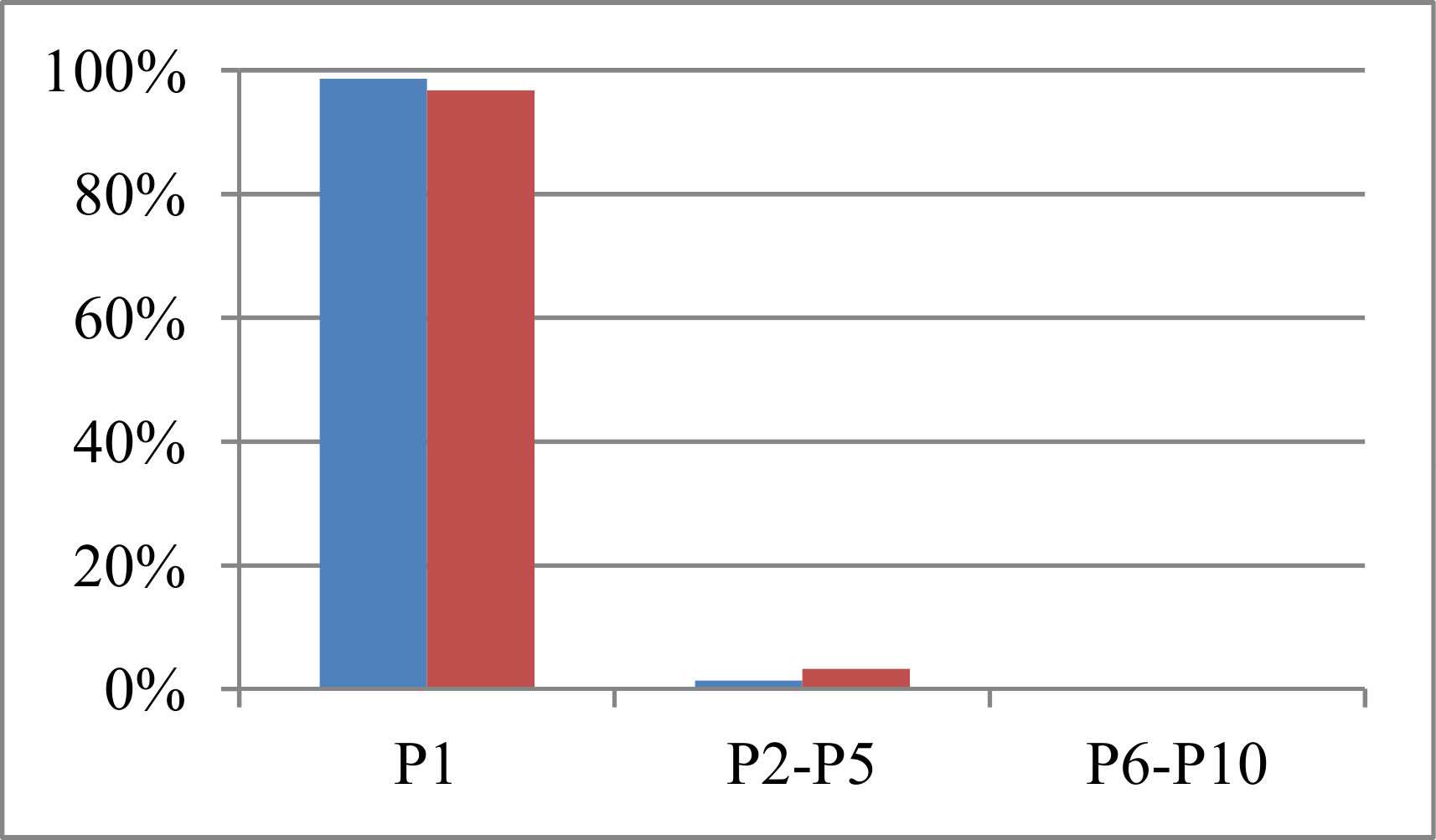}}
	\subfigure[\textsc{InPhO} Idea.]{\includegraphics[width=0.24\textwidth]{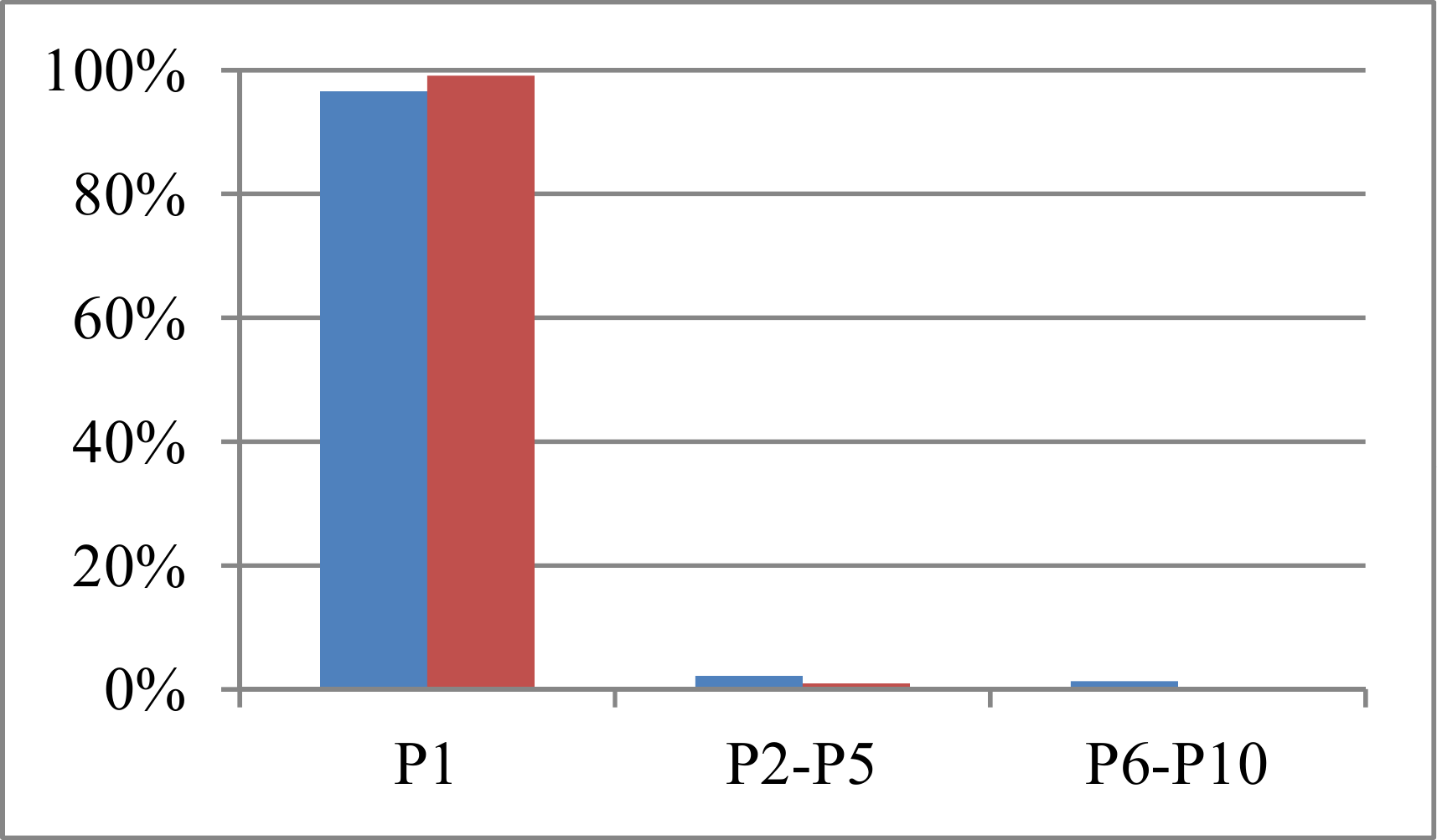}}
	\subfigure[\textsc{NELL}.]{\includegraphics[width=0.24\textwidth]{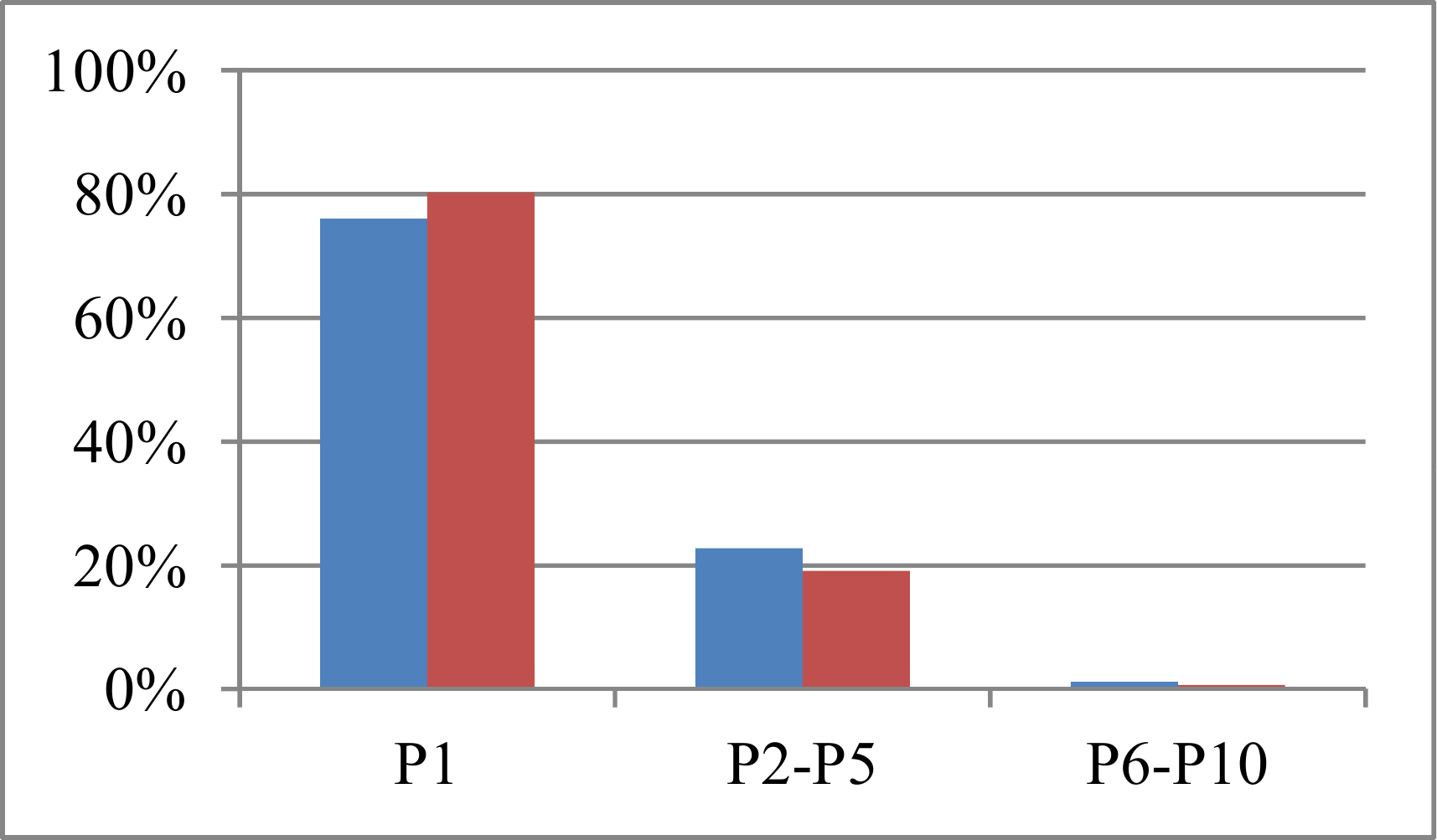}}
	\caption{Position of the correct matching candidates on a percentage basis if the entity has been under the top 10 search results (blue = \textsc{SPARQL}, red = \textsc{WikiStat}).}
	\label{fig:eval1_position}
\end{figure*}

Our experiments show that the \textsc{WikiStat} algorithm compared to the \textsc{SPARQL} algorithm with label and abstract has different strengths. While the \textsc{SPARQL} algorithm has been stronger on entities representing persons (here: \textit{Thinkers}) and documents (here: Journals), \textsc{WikiStat} achieved better results for philosophical concepts (here: Ideas) and the \textsc{NELL} dataset. The reason is that for persons and documents the naming is more accurate while philosophical concepts often have several possible equivalent names.% (e.g. ``Classicism''). 
Since the keywords that link to one specific \textsc{WikiPedia} entry cover multiple possible names while the abstract or the label often contains only one name, the \textsc{NELL} algorithm has a higher recall in these cases. Runtimes for the \textsc{WikiStat} and \textsc{SPARQL}-AL algorithm were comparable. 
If we compare the position of the correct matching candidates displayed in Figure~\ref{fig:eval1_position}, we observe that both algorithms were able to rank the correct candidates at the first position in over 97 \% of the cases for the \textsc{InPhO} and in over 77 \% of the cases for the \textsc{Nell} gold standard.

In all cases we obtain increased running times for the \textsc{Nell} gold standard, since \textsc{Nell} entities have often multiple labels.

\section{Conclusion and Future Work}

The aim of our Linked Humanities project is to enable non-technical humanities scholars to integrate and enrich local RDF repositories with the Linked Open Data cloud. To that end, we developed the linked open data enhancer (\textsc{LODE}) which provides intuitive user interfaces for linking and enhancing local RDF repositories while maintaining high quality collections. %The interfaces shield the Semantic Web specific algorithms for alignment candidate selection and property suggestions from the user.
The evaluation of two linking algorithms showed that they are able to provide high quality linking candidates with a response time of under $1.5$ second. We observed that the \textsc{SPARQL} algorithm performed better when linking persons and documents while \textsc{WikiStat} gained higher results for suggesting candidates for philosophical concepts and for the \textsc{NELL} benchmark.

In future work, we will perform a user study to evaluate the entire framework including the enhancement component. One aspect we aim to examine are comparisons between different (more sophisticated) context selection algorithms. Furthermore, we plan to add additional ``out of the box'' repositories apart from \textsc{DBPedia}. We are also continuously improving the linking and enhancement algorithms. Additionally, we plan to extend the enhancement interface so as to also allow the manual addition of novel content. 

\section{Acknowledgments}
This work has partly been funded by the DFG (NI 1364/1-1) and NEH (HG-50032-11). Many thanks to Kai Eckert for providing us with valuable feedback.

\bibliographystyle{abbrv}
\bibliography{lode} 

\end{document}